\documentclass[twocolumn,showpacs,preprintnumbers,amsmath,amssymb,pra,aps,superscriptaddress,floatfix]{revtex4-1}

\usepackage{graphicx}  
\usepackage{latexsym}

\begin{document}

\title{Strong spin-photon coupling in silicon}

\author{N.~Samkharadze$^*$}
\affiliation{QuTech and Kavli Institute of Nanoscience, Delft University of Technology, Lorentzweg 1, 2628 CJ Delft, The Netherlands}
\author{G.~Zheng$^*$}
\affiliation{QuTech and Kavli Institute of Nanoscience, Delft University of Technology, Lorentzweg 1, 2628 CJ Delft, The Netherlands}
\author{N.~Kalhor}
\affiliation{QuTech and Kavli Institute of Nanoscience, Delft University of Technology, Lorentzweg 1, 2628 CJ Delft, The Netherlands}
\author{D.~Brousse}
\affiliation{QuTech and Netherlands Organization for Applied Scientific Research (TNO), Stieltjesweg 1 2628 CK Delft, The Netherlands}
\author{A.~Sammak}
\affiliation{QuTech and Netherlands Organization for Applied Scientific Research (TNO), Stieltjesweg 1 2628 CK Delft, The Netherlands}
\author{U.~C.~Mendes}
\affiliation{Institut quantique and D\'{e}partment de Physique, Universit\'{e} de Sherbrooke, Sherbrooke, Qu\'{e}bec J1K 2R1, Canada}
\author{A.~Blais}
\affiliation{Institut quantique and D\'{e}partment de Physique, Universit\'{e} de Sherbrooke, Sherbrooke, Qu\'{e}bec J1K 2R1, Canada}
\affiliation{Canadian Institute for Advanced Research, Toronto, ON, Canada}
\author{G.~Scappucci}
\affiliation{QuTech and Kavli Institute of Nanoscience, Delft University of Technology, Lorentzweg 1, 2628 CJ Delft, The Netherlands}
\author{L.~M.~K.~Vandersypen}
\affiliation{QuTech and Kavli Institute of Nanoscience, Delft University of Technology, Lorentzweg 1, 2628 CJ Delft, The Netherlands}

\date{\today}

\begin{abstract}

We report the strong coupling of a single electron spin and a single microwave photon. The electron spin is trapped in a silicon double quantum dot and the microwave photon is stored in an on-chip high-impedance superconducting resonator. The electric field component of the cavity photon couples directly to the charge dipole of the electron in the double dot, and indirectly to the electron spin, through a strong local magnetic field gradient from a nearby micromagnet. This result opens the way to the realization of large networks of quantum dot based spin qubit registers, removing a major roadblock to scalable quantum computing with spin qubits.
\end{abstract}

\maketitle

Light-matter interaction has had profound impact on the development of quantum theory starting from the discovery of the photo-electric effect~\cite{Einstein05}: one single photon can release one single electron from a solid  provided the photon energy exceeds the electron binding energy of the material. This observation demonstrates that light consists of quanta, but does not rely on a coherent interaction between light and matter. In cavity quantum electrodynamics, a photon is stored in a cavity so that its interaction with a resonant atom or other two-level system in the cavity is enhanced to the point where a single quantum of energy is exchanged coherently between the cavity photon mode and the atom~\cite{Haroche06}. This regime is known as the strong-coupling regime and has been achieved across a wide range of experimental platforms, from atoms to superconducting qubits and self-assembled quantum dots, using either optical or microwave photons~\cite{Thompson92,Brune96,Wallraff04,Chiorescu04,Reithmaier04,Yoshie04}. Given that cavities extend over macroscopic distances, the coherent cavity-atom interaction can be used to indirectly couple well separated atoms coherently, offering a path to scalable quantum computing.

This prospect has motivated extensive theoretical and experimental work to achieve the strong-coupling regime with gate-defined semiconductor quantum dots, a leading platform for the realization of quantum circuits~\cite{Shulman12,Veldhorst15,Watson17,Zajac17}. Recently, strong coupling has been reported between a microwave photon and a charge qubit formed in a double quantum dot, an impressive achievement given the small electric dipole of a double dot and the short-lived charge qubit coherence~\cite{Mi16,Stockklauser17,Bruhat16}. Even more challenging, but also more desirable, is the strong coupling to a single electron spin~\cite{Mi17}. Compared to the electron charge, the electron spin has far superior coherence properties, but its direct interaction with the cavity magnetic field is exceedingly small~\cite{Haikka17}. Therefore, one must resort to indirect interaction of the electron spin to the cavity electric field by hybridization of the spin with the electron charge degree of freedom, without compromising spin coherence too severely in the process~\cite{Childress04,Burkard06,Trif08,Cottet10}. For a single spin, spin-charge hybridization can be achieved in a controlled way via a transverse magnetic field gradient~\cite{Pioro07,Kawakami14,Hu12,Viennot15,Beaudoin16,Benito17}.
 
Here we report the observation of vacuum  Rabi splitting of a single electron spin resonant with an on-chip microwave cavity, the telltale sign of strong coupling. We show how the spin-photon coupling strength is controlled by the charge qubit settings and extract all the relevant coupling strengths and decay rates. At a spin-photon coupling strength of 10 MHz, we observe cavity and spin decay rates of 4.1 and 1.8 MHz, respectively. 
 
\begin{figure}
 \includegraphics[width=\columnwidth]{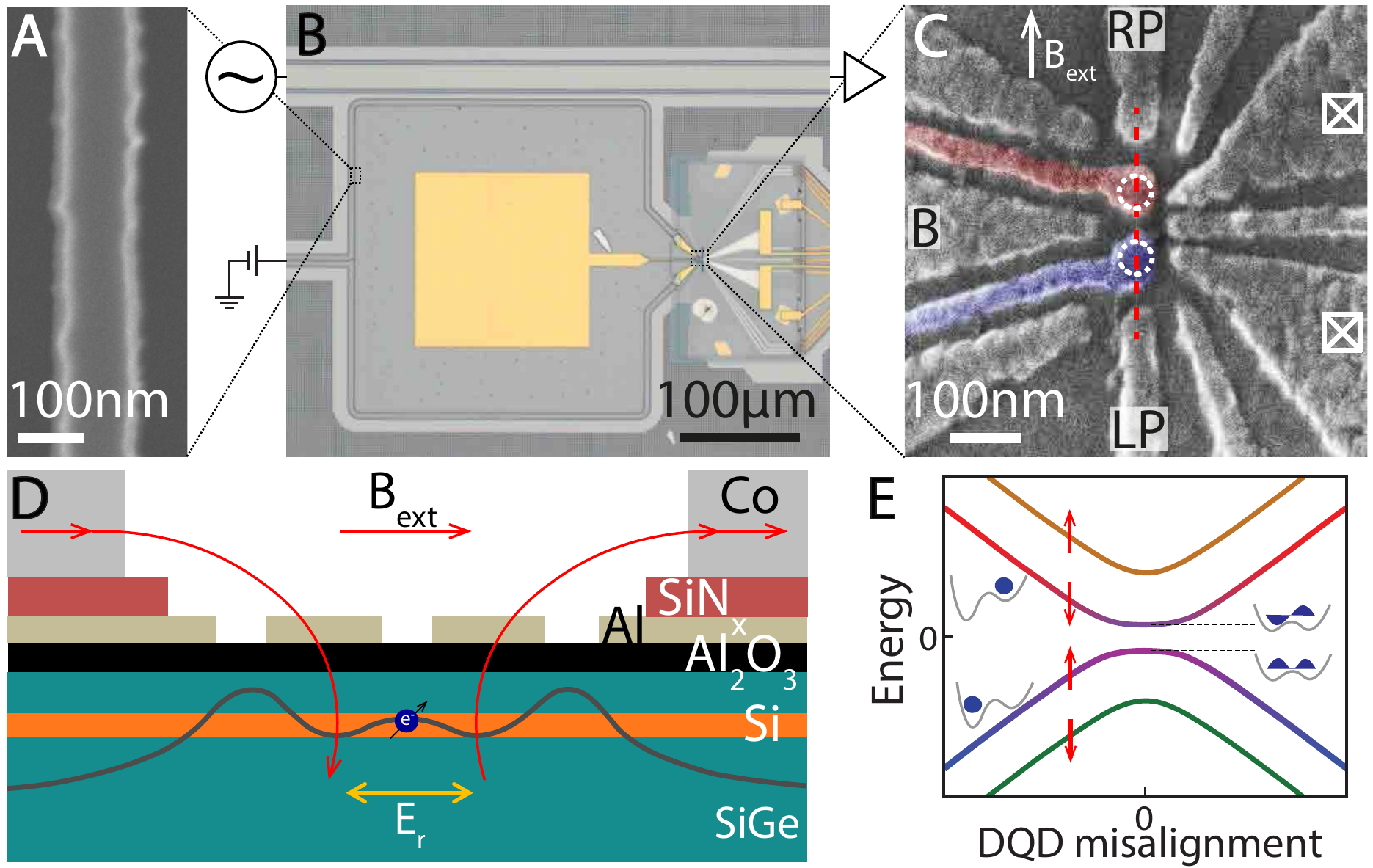}
 \caption{\label{f1}
Device images and schematic
(A) Scanning electron micrograph of a segment of the NbTiN resonator center conductor.
(B) Optical micrograph of the resonator (square shape) adjacent to the feed line (top) and double dot (right). The yellow square in the center is a bond pad to bias gate B.
(C) Scanning electron micrograph showing the gates used to form the double quantum dot (white dotted circles indicate dot positions). The purple and red colored gates are connected to the resonator ends.
(D) Schematic cross-section of the quantum dot along the red dashed line in panel (d), showing the Si quantum well with SiGe buffer and barrier layers, and the Al$_2$O$_3$ and SiN$_x$ dielectrics separating the substrate from the Al gates and Co micromagnets. In the experiment, a single electron moves in the double dot potential landscape (grey line) in response to the resonator electric field, $E_{r}$. A magnetic field is applied in the plane of the quantum well. The Co micromagnets create an additional magnetic field component, with a different orientation between the two dots.
(E) The DQD energy levels as a function of DQD misalignment. Near $\epsilon =0$, the left and right dot levels hybridize, forming bonding and anti-bonding states that define a charge qubit~\cite{Hayashi03}. Each of the DQD levels is split by the Zeeman energy. The micromagnet causes spin and orbital levels to hybridize as well.
}
\end{figure}

Figure~\ref{f1} shows device images and a device schematic (see also Fig.~S1). The superconducting cavity consists of a  NbTiN coplanar resonator with a narrow center conductor and remote ground planes (Fig.~\ref{f1}A,B), capacitively coupled to a feed line. The cavity resonator is wrapped in a square shape and its two ends are connected to two Al gates that extend over the quantum dot locations. The resonator materials choice and dimensions give it a high characteristic impedance of about 1 k$\Omega$ that enhances the coupling $g_c$ to the double dot charge dipole~\cite{Samkharadze16,Stockklauser17}, and make it resilient to in-plane magnetic fields of over 6 T~\cite{Samkharadze16}. The double quantum dot (DQD) is formed electrostatically in an undoped Si/SiGe quantum well (natural isotopic abundance), using a single layer of Al gates (Fig.~\ref{f1}C). A positive bias on a gate accumulates electrons in the quantum well underneath, a negative bias repels electrons (Fig.~S1D). An in-plane magnetic field $B_{\mathrm{ext}}$ induces a Zeeman splitting on an electron in the DQD. Two Cobalt micromagnets placed near the quantum dots (Fig.~S1B,C) produce a local gradient in the static magnetic field. As a result, when an electron oscillates between the two dots, it experiences an oscillating transverse magnetic field, providing the necessary (indirect) spin-charge hybridization that allows an electric field to couple to the spin~\cite{Pioro07,Kawakami14,Viennot15} (Fig.~\ref{f1}E).

\begin{figure}
 \includegraphics[width=\columnwidth]{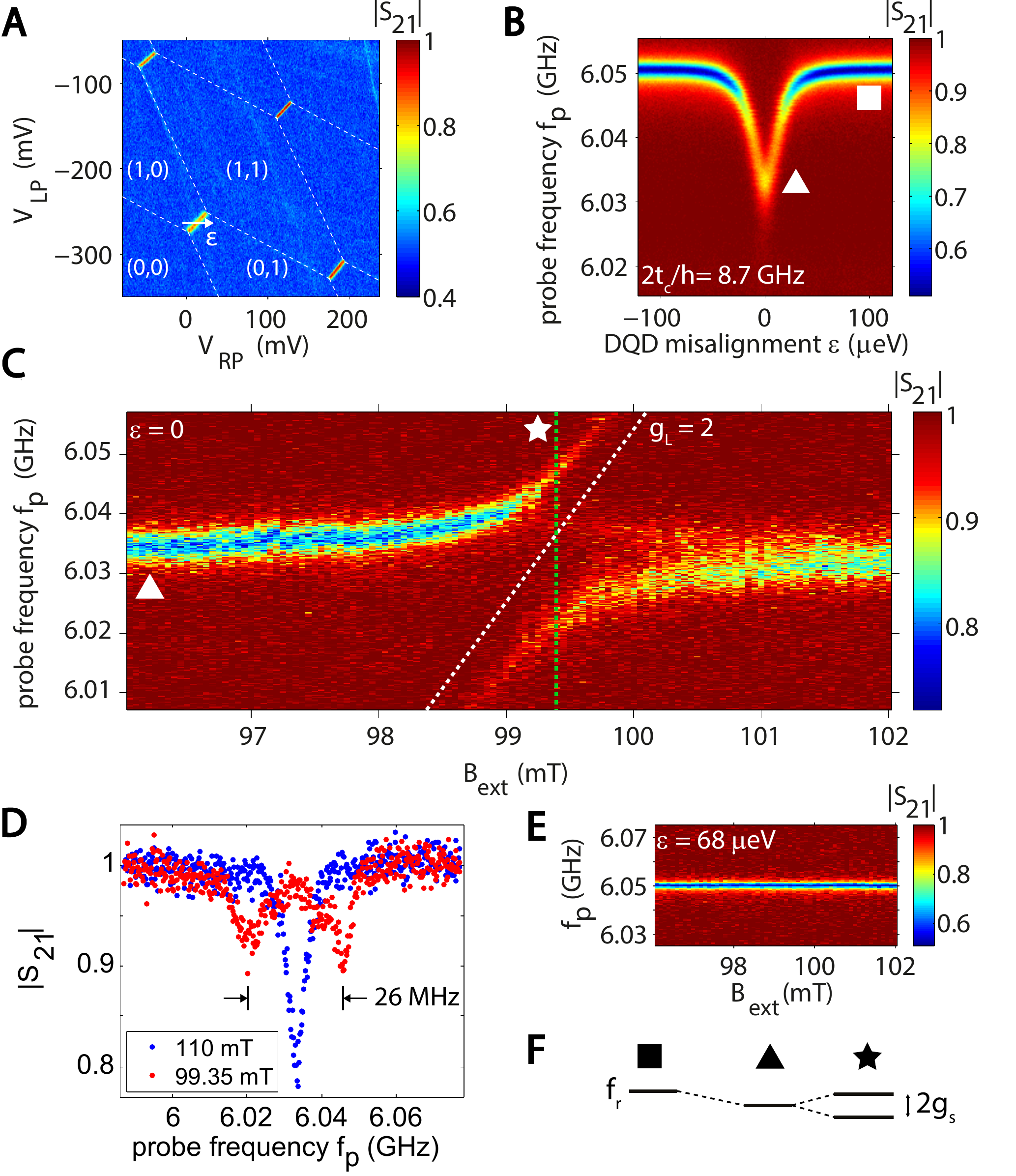}
 \caption{\label{f2}
Strong spin-photon coupling.
 (A) Transmission as a function of two gate voltages that control the potential of the two dots. At the four bright lines, the electron can move between the dots. The dashed lines connecting the short lines indicate alignment of a dot with a reservoir electrochemical potential. Labels indicate the electron number in the two dots.
 (B) Transmission as a function of $\epsilon$ (along the full white line in panel A) and $f_p$. At large $|\epsilon|$, we measure the bare resonator transmission (square symbol). Near $\epsilon = 0$, the DQD charge qubit interacts dispersively with the cavity frequency, leading to a characteristic frequency shift (triangle symbol).
 (C) Transmission as a function of $B_{\mathrm{ext}}$ and $f_p$. When $B_{\mathrm{ext}}$ makes the spin spitting resonant with the resonator frequency (star symbol), a clear avoided crossing occurs, which we attribute to the strong coupling of a single spin and a single photon. The dotted line shows the expected spin splitting for a spin in silicon. 
 (D) Line cut through panel C at the position of the green vertical line (red data points) and line cut at 110 mT (blue points). The red data shows clear vacuum Rabi splitting.
 (E) Similar to C but with the DQD misaligned, so the electron cannot move between the two dots. The spin-photon coupling is no longer visible.
 (F) Schematic representation of the transmission resonance of the superconducting cavity. The bare transmission resonance (square) is shifted dispersively by its interaction with the charge qubit (triangle), and splits when it is resonant with the spin qubit (star).
}
\end{figure}

We apply a probe tone to the feed line at frequency $f_p$ and record the transmission through the feed line (unless indicated, all transmission plots show the normalised amplitude of the transmission through the feed line). With the DQD tuned to keep the electron fixed in one of the dots, the transmission shows a dip for $f_p$ near 6.05 GHz, the bare resonance frequency $f_r$ of the NbTiN resonator (Fig.~\ref{f2}B square symbol). From the linewidth, we find the bare resonator decay rate $\kappa_r/2\pi = 2.7$ MHz, with an internal loss rate $\kappa_{int}/2\pi = 1.5$ MHz. In what follows, we monitor the transmission through the feed line at low probe power (below -125 dBm, corresponding to $< 1$ photon in the resonator) to tune up the DQD, characterize the charge-photon interaction, and study spin-photon coupling.

To characterize the charge-photon interaction, we tune the DQD to a regime where the electron can move back and forth between the two dots in response to the cavity electric field. Such motion is possible whenever the electrochemical potentials of the two dots are aligned, i.e. where it costs equal energy for an electron to be in either dot. This occurs for specific combinations of gate voltages, seen as the short bright lines in Fig.~\ref{f2}A, where the charge-photon interaction modifies the transmission~\cite{Frey12}. We focus on the lower left line, which corresponds to the last electron in the DQD.

In order to place the charge-photon interaction in the dispersive regime, we set the charge qubit splitting $f_c$ in the range of 8 to 15 GHz, so that $f_c$ is always well above $f_r$. We measure $f_c$ using two-tone spectroscopy, as detailed below. In the dispersive regime, the charge-photon interaction results in a frequency shift of the resonator (Fig.~\ref{f2}F). In Fig.~\ref{f2}B, we observe the characteristic dependence of this dispersive shift on the DQD misalignment $\epsilon$. At $\epsilon = 0$, the electron can most easily move between the dots, hence the electrical susceptibility is the highest and the dispersive shift the largest (triangle). At $\epsilon = 0$, the magnitude of the dispersive shift is approximated by $(g_c/2\pi)^2/(f_c - f_r)$, where the charge-photon coupling strength $g_c$ is mostly fixed by design and the detuning between $f_c$ and $f_r$ can be adjusted. From a fit based on input-output theory~\cite{Peterson12}, we extract a charge-photon coupling strength $g_c/2\pi$ of $\sim 200$ MHz.

\begin{figure}
	\includegraphics[width=\columnwidth]{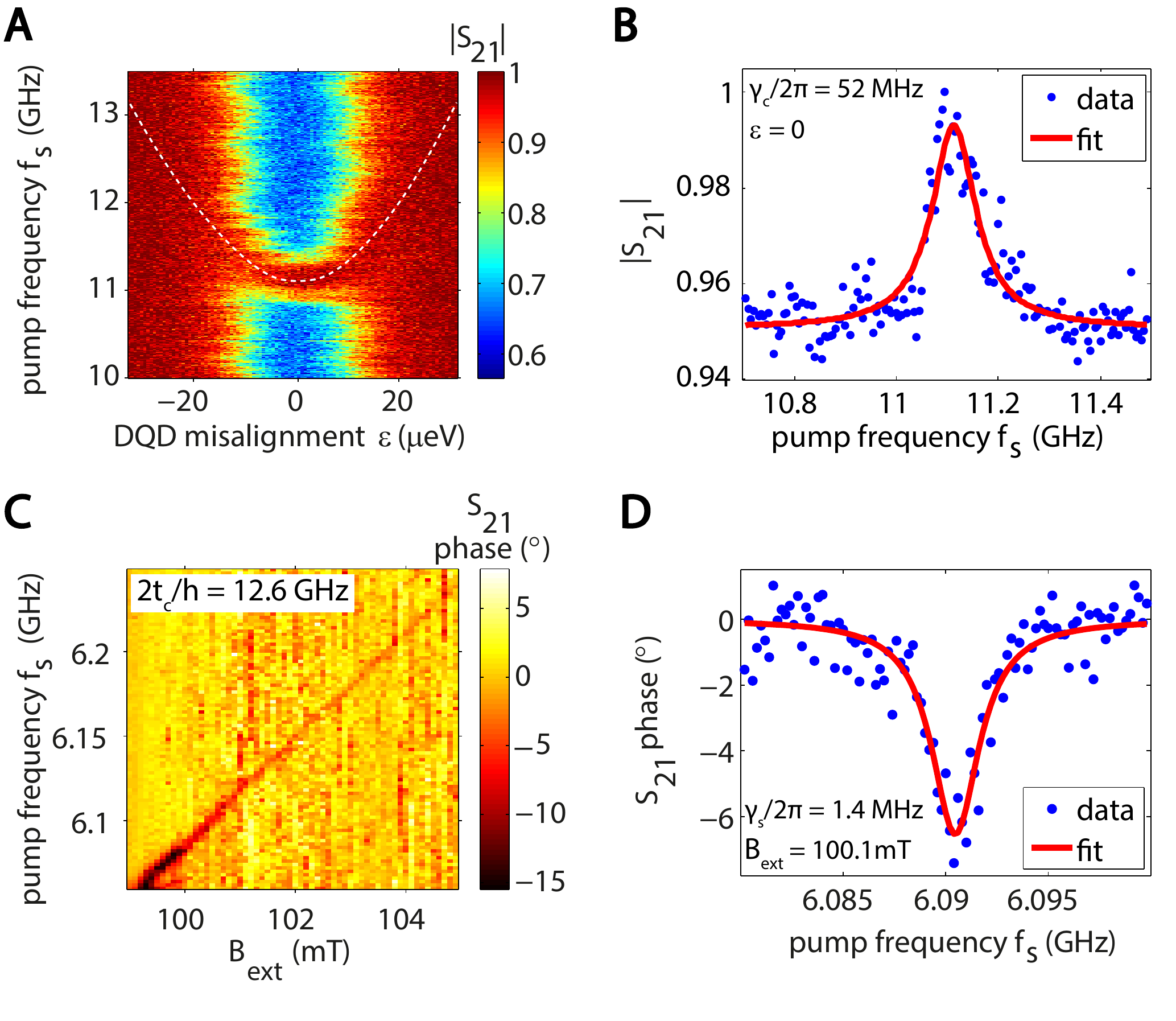}
	\caption{\label{f4}
		Two-tone spectroscopy of the charge and spin qubit
		(A) Transmission at $f_p = 6.041$ GHz as a function of DQD misalignment $\epsilon$ and the frequency of a second tone (pump frequency) that is applied to gate LP. When the second tone is in resonance with the charge qubit splitting (white dotted line), the steady-state occupation of the charge qubit is changed, and due to the charge-photon coupling, this is reflected in a modified dispersive shift of the resonator. 
		(B) Line cut at $\epsilon=0$, from which we extract a charge qubit dephasing rate of 52 MHz.
		(C) Transmission (phase response) at $f_p = 6.043$ GHz as a function of $B_{\mathrm{ext}}$ and the pump frequency applied to gate LP. When the pump frequency is in resonance with the spin qubit splitting, the steady-state occupation of the spin qubit is changed, and due to the spin-photon coupling, this is reflected in a modified response of the resonator. The slope of the response corresponds to a spin with $g_L = 2$, as expected.
		(D) Line cut at $B_{\mathrm{ext}} = 100.1$ mT, from which we extract a spin qubit dephasing rate of 1.4 MHz. 
	}
\end{figure}

To probe coherent spin-photon coupling, we keep the charge sector parameters constant so that the interaction with charge remains dispersive. By varying $B_{\mathrm{ext}}$, we control the spin splitting such that the interaction with the spin goes from dispersive to resonant. On resonance, spin and photon hybridize (Fig.~\ref{f2}F triangle).
Experimentally, we record the transmission through the feed line as a function of the strength of an in-plane magnetic field $B_{\mathrm{ext}}$ (the total field is the vector sum of external field and the micromagnet stray field) and the probe frequency $f_p$ applied to the feed line. As expected, the cavity resonance seen in transmission is (nearly) independent of $B_{\mathrm{ext}}$ at large spin-resonator detuning. When the spin splitting approaches resonance with the resonator frequency, we observe a strong response in the form of an anti-crossing (Fig.~\ref{f2}C). The slope $f/B_{\mathrm{ext}}$ of the slanted branch corresponds to $g_L \mu_B$, with $\mu_B$ the Bohr magneton and $g_L\approx 2$ the Land\'e $g$-factor of an electron spin in Si. The observed avoided crossing is thus a clear signature of the coherent hybridization of the spin qubit with a single microwave photon.

The line cut, indicated by the dashed green line in Fig.~\ref{f2}C and shown in Fig.~\ref{f2}D, reveals two well separated peaks. This feature is known as the vacuum Rabi splitting and is the hallmark of strong coherent spin-photon coupling. The peak separation is about 26 MHz, corresponding to a spin-photon coupling strength $g_s/2\pi$ of 13 MHz. 
The cavity decay rate can be extracted independently from the linewidth away from spin-photon resonance, here $\kappa/2\pi = 5.4$ MHz (here the cavity dispersively interacts with the charge, so $\kappa > \kappa_r$~\cite{Frey12}). The spin dephasing rate $\gamma_s/2\pi = 2.5$ MHz is independently obtained from two-tone spectroscopy of the spin transition (discussed next). We observe that $g_s > \kappa, \gamma_s$, satisfying the condition for strong coupling of a single electron spin to a single microwave photon.

Two-tone spectroscopy of the charge and spin qubits allows us to independently extract the respective qubit splittings and dephasing rates. In Fig.~\ref{f4}A,B the second tone is resonant with the charge qubit splitting around 11.1 GHz, with a dependence on $\epsilon$ described by $h f_c=\sqrt{4t_c^2+\epsilon^2}$, with $t_c$ the interdot tunnel coupling and $h$ Planck's constant, see the white dotted line (neglecting spin-charge hybridization). In this case, we extract from the linewidth a charge qubit dephasing rate $\gamma_c/2\pi$ of 52 MHz. In Fig.~\ref{f4}C,D, we sweep the second tone through the spin resonance condition while keeping the spin-photon system in the dispersive regime. We observe a linear dependence of spin splitting on $B_{\mathrm{ext}}$, with a slope corresponding to $g_L = 2$, as expected. At $2t_c/h=12.6$ GHz, we extract $\gamma_s/2\pi = 1.4$ MHz from the linewidth. 
This is somewhat larger than the $\sim 0.3$ MHz single-spin dephasing rates observed in a single Si/SiGe quantum dot~\cite{Kawakami14,Watson17,Zajac17}, as can be expected given that an electron in a DQD at $\epsilon=0$ is more susceptible to charge noise, which affects spin coherence through the magnetic field gradient \cite{Hu12,Beaudoin16,Benito17}.

\begin{figure}
	\includegraphics[width=\columnwidth]{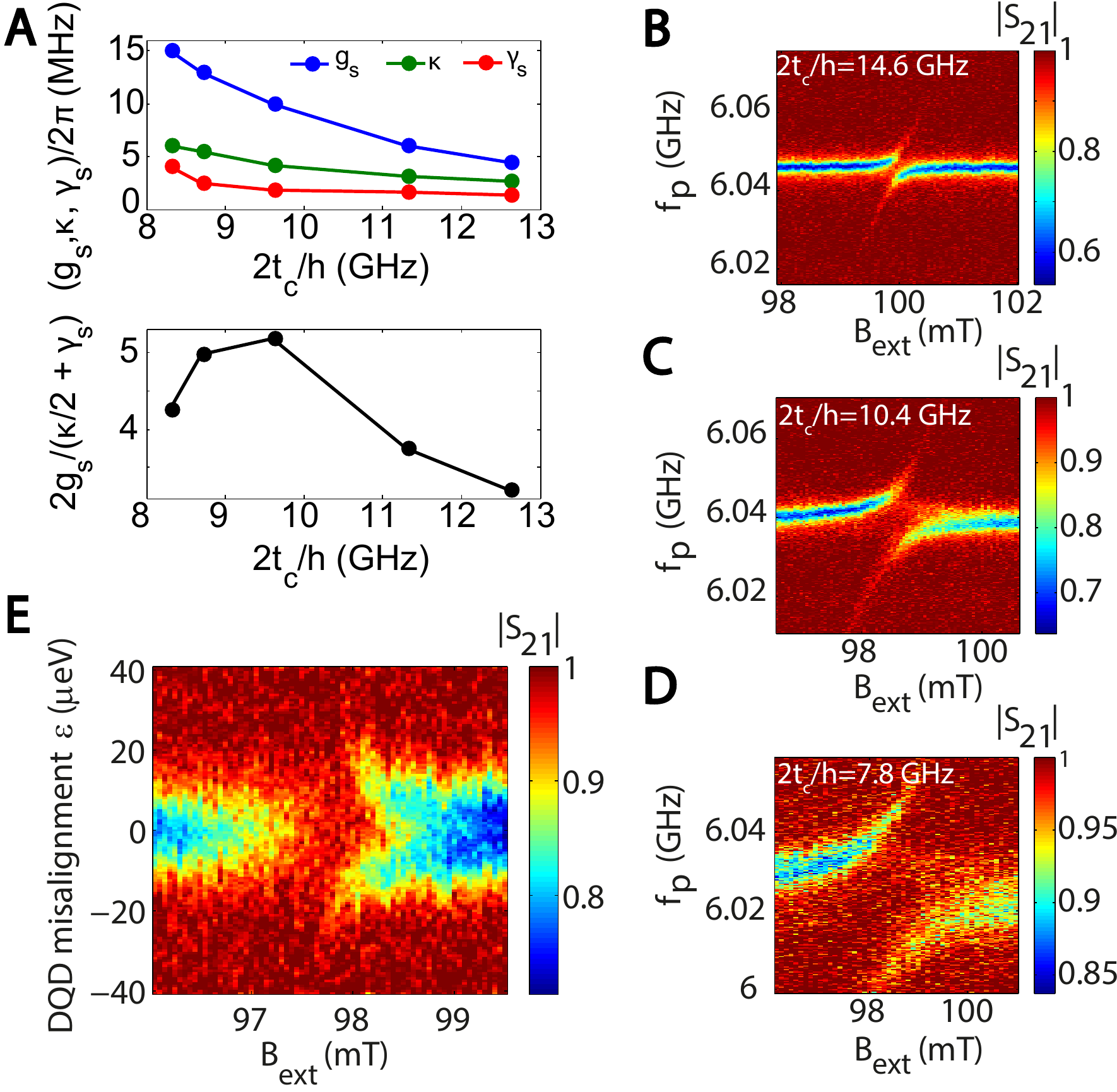}
	\caption{\label{f3}
Control of the spin-photon coupling.
(A) The dependence on DQD tunnel coupling of $g_s$, $\kappa$, $\gamma_s$ (upper panel) and the ratio of peak splitting to linewidth $2g_s/(\gamma_s+\kappa/2)$ (lower panel). While all three separate quantities increase with lower $2t_c$, the ratio $2g_s/(\gamma_s+\kappa/2)$, which is the most relevant quantity, shows an optimum value around $f_c = 10$ GHz.
(B-D) Similar data to Fig.~\ref{f2}C for three different values of DQD tunnel coupling, as indicated.
(E)  Transmission as a function of $B_{\mathrm{ext}}$ and $\epsilon$. Where the blue band is interrupted, the Zeeman splitting is resonant with the (dispersively shifted) resonator.		
	}
\end{figure}

The spin-photon hybridization can be controlled with gate voltages. Indeed, by moving away from $\epsilon = 0$, the photon and charge no longer hybridize, and then also the spin-photon coupling vanishes, as expected (Fig.~\ref{f2}E). Furthermore, at $\epsilon=0$ the spin-photon coupling strength can be approximated as $g_s = g_c \Delta B_x/(2 t_c-f_r)$ (provided the magnetic field profile is symmetric relative to the DQD)~\cite{Hu12,Beaudoin16,Benito17}. Here $\Delta B_x$ is the difference in the transverse field between the two dots. Starting from large $t_c$, reducing $t_c$ increases charge-photon admixing, and thus indirectly spin-photon coupling as well, as seen experimentally in Figs.~\ref{f3}B-D. With increased charge-photon admixing, the asymmetry in the intensity of the two branches also increases, as expected in this system composed of photon, charge and spin~\cite{Benito17}, and an additional feature appears close to the lower branch (discussed in the Supplementary Information). The variation of $g_s$ with $t_c$ is summarised in Fig.~\ref{f3}A. However, as seen in the same figure, with lower $t_c$ the spin decay rate $\gamma_s$ increases as well, as does the cavity decay rate $\kappa$~\cite{Benito17}.  Ultimately, we wish to maximize the peak separation over linewidth, $2g_s/(\gamma_s+\kappa/2)$. In this respect, there is an optimal choice of tunnel coupling, as seen from Fig.~\ref{f3}A.

Finally, we study how close together the charge and spin sweet spots occur, where the relevant frequency (charge or spin) is to first order insensitive to the DQD misalignment.
The charge sweet spot is seen in Fig.~\ref{f2}B, at $\epsilon = 0$ and $f_p = 6.03$ GHz. If the micromagnets are placed symmetrically with respect to the DQD (as in Fig.~\ref{f1}D), the total magnetic field magnitude is symmetric around the center of the DQD. In this case, the spin splitting has no first order dependence on $\epsilon$ at $\epsilon = 0$ and the charge and spin sweet spots coincide. For asymmetrically placed magnets, the spin sweet spot occurs away from $\epsilon = 0$. To find the spin sweet spot, we vary $\epsilon$ and $B_{\mathrm{ext}}$ at $f_p = 6.03$ GHz (Fig.~\ref{f3}E). Throughout the blue band, $f_p$ is resonant with the cavity frequency (in the dispersive charge-photon regime). Where the blue band is interrupted, the magnetic field brings the spin on resonance with the cavity photon, spin and photon hybridize, and the transmission is modified. As expected, we see that this spin-photon resonance condition slightly shifts in magnetic field as a function $\epsilon$~\cite{Viennot15}. The value of $\epsilon$ where this shift has no first order dependence on $\epsilon$ occurs close to $\epsilon = 0$, i.e. the spin sweet spot lies close to the charge sweet spot.

The strong coupling of spin and photon not only opens up a new range of physics experiments but is also the crucial requirement for coupling spin qubits at a distance via a superconducting resonator. Given the large dimensions of the resonators compared to the double dot dimensions, multiple spin qubits can interact to and via the same resonator, enabling scalable networks of interconnected spin qubit registers~\cite{Vandersypen17}. Importantly, the spin-photon coupling can be switched on or off on nanosecond timescales using gate voltage pulses that control the double dot misalignment and tunnel coupling, facilitating on-demand coupling of one or more spins to a common resonator.\\

$^*$ These authors contributed equally to this work.

\section*{Acknowledgments}
We thank J. Taylor, P. Scarlino, A. Yacoby, J. Kroll and members of the spin qubit team at QuTech for useful discussions, and L. Kouwenhoven and his team for access to NbTiN films. This research was undertaken thanks in part to funding from the European Research Council (ERC Synergy Quantum Computer Lab), the Dutch Foundation for Scientific Research (NWO through the Casimir Research School), Intel Corporation,  the Canada First Research Excellence Fund and NSERC.

\end{document}